\documentclass[twocolumn,showpacs,amsmath,amsfonts,amssymb,superscriptaddress,nofootinbib,preprintnumbers]{revtex4-1}
\pdfoutput=1
\usepackage[usenames,dvipsnames]{color}
\usepackage[colorlinks=true, linkcolor=BrickRed, citecolor=Blue, urlcolor=Blue, filecolor=Blue]{hyperref}
\usepackage{longtable,epsfig,graphicx,verbatim,xspace,multirow,ulem}
\usepackage{grffile}

\begin{document}
\title{Maximum amplitude of the high-redshift 21-cm absorption feature}

\author{Pablo Villanueva-Domingo}
\email{pablo.villanueva@ific.uv.es}
\author{Olga Mena}
\email{olga.mena@ific.uv.es}
\affiliation{Instituto de F\'isica Corpuscular (IFIC),
  CSIC-Universitat de Valencia,\\
Apartado de Correos 22085,  E-46071, Spain}
\author{Jordi Miralda-Escud\'e}
\email{miralda@icc.ub.edu}
\affiliation{Institut de Ci\`encies del Cosmos, Universitat de Barcelona (IEEC-UB), Barcelona, Spain}
\affiliation{Instituci\'o Catalana de Recerca i Estudis Avan\c cats,
 Barcelona, Spain}

\begin{abstract}
We examine the maximum possible strength of the global 21-cm absorption
dip on the Cosmic Background Radiation at high-redshift caused by the
atomic intergalactic medium, when the Lyman-$\alpha$ coupling is maximum,
assuming no exotic cooling mechanisms from interactions with dark matter.
This maximum absorption is limited by three inevitable factors that need
to be accounted for: \textit{(a)} heating by energy transferred
from the Cosmic Background Radiation to the
hydrogen atoms via 21-cm transitions, dubbed as \textit{21-cm heating};
\textit{(b)} Ly$\alpha$ heating by scatterings of Ly$\alpha$ photons
from the first stars; \textit{(c)} the impact of the expected density fluctuations in the intergalactic gas in standard
Cold Dark Matter theory, which reduces the mean 21-cm absorption signal.
Inclusion of this third novel effect reduces the maximum global 21-cm
absorption by $\sim 10\%$. Overall, the three effects studied here reduce the
21-cm global absorption by $\sim 20\%$ at $z \simeq 17$.

\end{abstract}

\maketitle


\section{Introduction}

The first non-linear structures containing baryonic matter in the
Universe are expected to form in the Cold Dark Matter (CDM) model of
structure formation at redshift $z\sim 20$, when the intergalactic
medium (IGM) was mostly atomic and had been adiabatically cooling since the
residual electrons left over from recombination could no longer keep
the temperature equal to that of the Cosmic Background Radiation (CBR).
The resulting temperature was $\overline{T}_{ad} \simeq 8\, {\rm K}
[(1+z)/20]^2$ (e.g. \cite{Pritchard:2011xb}), which
implies a comoving Jeans scale for the atomic gas (with adiabatic sound speed $c_s^2=5k_B\overline{T}_{ad}/(3\mu m_p)$, where $\mu$ is the mean molecular weight and $m_p$ the mass of the proton) of
\begin{equation}
\lambda_{J} = \frac{\pi (1+z)}{ H(z) } \sqrt{\frac{40 k_B\overline{T}_{ad}}{9\mu m_p}} \simeq
 9.1 \, \left( {1+z\over 20} \right)^{1/2} \, {\rm kpc} ~,
\label{jeans}
\end{equation}
where $H(z)$ is the Hubble expansion rate. Throughout this paper we use the standard $\Lambda$CDM model with $H_0= 68\, {\rm km}\,{\rm s}^{-1}\,{\rm Mpc}^{-1}$, $\Omega_{m0}=0.308$, and $\Omega_{b0}=0.048$, see e.g.~\cite{Ade:2015xua}.

  There is a strong interest in cosmology in detecting density
fluctuations of the IGM at this early epoch, because
they are probing the smallest scales of primordial fluctuations that
can be detected, and the conditions that gave rise to the formation of
the first stars in the Universe, (e.g. \citep{Munoz:2019hjh}). Absorption of the CBR radiation in
the redshifted 21-cm hyperfine line of atomic hydrogen offers the most
plausible method to detect this primitive and coldest state of baryonic
matter.

 Experiments targeting a measurement of the cosmological 21-cm line,
arising from spin-flip transitions between the triplet and the ground
singlet states in neutral  hydrogen in the IGM, open a
new window to the epochs of reionization and the dark age of the
Universe. In particular, observation of 21-cm absorption at high redshift probes the epoch
when light from the first stars, which had not yet ionized or heated the IGM, was able to
couple the spin and kinetic temperatures of the atomic hydrogen through the
Wouthuysen-Field effect \cite{Wouthuysen:1952,Field:1958}. Although initial studies
suggested that 21-cm absorption might be swamped by rapid heating of the IGM, and the
corresponding 21-cm emission, by
the same Ly$\alpha$ photons responsible for the spin and kinetic temperature couplings
(the Ly$\alpha$ heating effect \cite{Madau:1996cs}), this heating was shown
to be small and to allow for an extended epoch of 21-cm absorption if heating caused by
X-ray emission of the first stars is not very strong
 \cite{Chen:2003gc,Hirata:2005mz,Furlanetto:2006fs,Venumadhav:2018uwn}.
This offers promising prospects for 21-cm observations of the primitive, cold atomic medium, as a
probe to primordial mass fluctuations at very small scales.

 A number of observatories aim to detect the 21-cm cosmological line.
An early result by The Experiment to Detect the Global Epoch of reionization
Signature (EDGES) \cite{Bowman:2012hf} claimed a 21-cm absorption temperature
at a redshift $z\sim 17$ twice larger than the maximum allowed
one in the absence of any heating, and for the baryon density derived
from CBR fluctuations and primordial
nucleosynthesis \cite{Bowman:2018yin}. Many studies explored exotic
scenarios of strong interactions of baryons with dark matter that
might further cool the early atomic IGM \cite{Liu:2018uzy,Kovetz:2018zan,Barkana:2018lgd}, but the EDGES result is likely affected by foregrounds or systematics \cite{Hills:2018vyr,Bradley:2018eev,Sims:2019kro}. 
Other high-redshift 21-cm searches include the future Large-Aperture Experiment to Detect the Dark Ages (LEDA) \cite{Greenhill:2012mn}, and the Moon orbiting space observatory Dark Ages RadioExperiment (DARE) \cite{Burns:2011wf}.

  To help interpret future results of the 21-cm cosmological signal, it is useful to derive the maximum possible global 21-cm absorption in standard cosmology, taking into account and accurately including any effects that may impact
the average amplitude of the absorption. This is the goal of the present study, which examines relevant astrophysical effects at the epoch of interest that may substantially modify the global 21-cm absorption dip.
Although many calculations and models have been done (see \textit{e.g.} \cite{Xu:2018shq} for an analytical estimation taking into account inhomogeneous distributions of the gas), we include here
a comprehensive study of the global effect of baryon density
fluctuations, whose impact has not been analyzed from an analytical perspective in the literature.

The structure of the paper is as follows. We review in Sec.~\ref{sec:sec1} the basics of the 21-cm global signal, discussing the imprint of Ly$\alpha$ emission and heating sources. Section~\ref{sec:sec2} describes the effect of IGM density fluctuations on the global signal. Results are presented in Sec.~\ref{sec:sec3}, accounting separately for each heating contribution first, and then illustrating the final 21-cm absorption amplitude including all effects. We conclude in Sec.~\ref{sec:sec4}.

\section{The 21-cm signal}
\label{sec:sec1}

 The 21-cm emission and absorption signal arises from the population of the
singlet and triplet states of the atomic hydrogen hyperfine structure,
$n_0$ and $n_1$ respectively, governed by the spin temperature $T_S$ through
the relation $n_1/n_0=3 \; \exp(-T_*/T_S)$, with $T_*=h\nu_0=0.0682$ K.
The differential brightness temperature in 21-cm absorption can be written as \cite{Pritchard:2011xb,Furlanetto:2006jb}
\begin{equation}
\delta T_b(\Delta) = T_0 \Delta \left(1- \frac{T_\gamma}{T_S(\Delta)} \right)~,
\label{eq:dTbdelta}
\end{equation}
where $T_\gamma$ is the CMB photon temperature,
$\Delta =1+\delta = \rho/\bar{\rho}$ is the gas density divided by its
mean value, $T_0=27 \, x_{\rm HI} \, \sqrt{(1+z)/10}$~mK is the global emission
temperature at the mean density for the cosmological model we use when
$T_S \gg T_\gamma$, and $x_{\rm HI}$ the neutral hydrogen
fraction.\footnote{The 21-cm signal is also affected by peculiar velocity
gradients, but in the optically thin regime the mean value is not affected
because peculiar velocities only redistribute the absorption over frequency.}

In this work we focus on the epoch prior to reionization, setting
$x_{\rm HI} \simeq 1$. The spin temperature $T_S$ evolution is determined
by various energy exchange processes: \textit{i)} absorption and emission
of CMB photons, coupling $T_S$ to $T_\gamma$;
\textit{ii)} collisions with electrons, protons or other H atoms, coupling
$T_S$ to the gas kinetic temperature $T_K$; and
\textit{iii)} the Wouthuysen-Field mechanism \cite{Wouthuysen:1952,Field:1958},
\textit{i.e.}, Ly$\alpha$ photon scattering, coupling $T_S$ to the
\textit{color} radiation temperature $T_\alpha$, which measures the slope
of the radiation background spectrum near the center of the Ly$\alpha$ line
where most of the scatterings occur. The spin temperature is rapidly set to
an equilibrium between these processes determined by the equation
\begin{equation}
T_S^{-1} = \frac{T_\gamma^{-1} + x_c T_K^{-1} + x_\alpha T_\alpha^{-1}}{1+ x_c + x_\alpha}~,
\label{eq:Tspin}
\end{equation}
where $x_c$ and $x_\alpha$ are coupling coefficients for collisions and
Ly$\alpha$ scattering, respectively. In general, we can safely assume that
$T_\alpha \simeq T_K$~\cite{Pritchard:2011xb,Furlanetto:2006jb}. We make use of the fits from \cite{Furlanetto:2006su,Zygelman_2005} for $x_c$, although coupling by collisions is important only at high redshifts. The coupling after star formation has started is driven
therefore by Ly$\alpha$ scattering,
\begin{equation}
x_\alpha = \frac{16 \pi^2 T_* e^2 f_{12}}{27 A_{10} T_\gamma m_e c}S_\alpha J_\alpha~,
\end{equation}
where $f_{12}$ is the oscillator strength of the Ly$\alpha$ transition, $A_{10}$ the spontaneous decay rate,
$m_e$ the electron mass of the electron, $e$ the charge of the electron, $c$ the speed of light,
$S_\alpha$ is an order unity correction factor accounting for the detailed
shape of the spectrum near the resonance, and $J_\alpha$ is the photon
Ly$\alpha$ flux per unit frequency at the Ly$\alpha$ line center. We
adopt the wing approximation for $S_\alpha$ obtained in
\cite{Furlanetto:2006fs}. We now explain how the evolution of the two crucial
quantities for 21-cm absorption, the Ly$\alpha$ flux $J_\alpha$ and the
kinetic temperature $T_K$, are obtained.

\subsection{Ly$\alpha$ flux}
\label{sec:appb}

The Ly$\alpha$ flux from direct stellar emission of UV photons,
$J_{\alpha\star}$, is given by the sum over the Ly-$n$
levels which can lead to a $2p \rightarrow 1s$ transition through a
decaying cascade. Photons redshifting to a Lyman resonance are always
absorbed owing to the high optical depth of the IGM. Photon reaching
the Ly-$n$ resonance at redshift $z$ have to be
emitted at a redshift below $1+z_{max,n} =
 (1+z)\,[1-(n+1)^{-2}]/[1-n^{-2}]$. Defining the probability to
generate a Ly$\alpha$ photon after absorption by the $n$ level as
recycled fraction of level $n$, $f_{rec}(n)$
\citep{Pritchard:2005an}, the total Ly$\alpha$ flux can be written in
terms of the photon comoving emissivity $\epsilon_\alpha(\nu,z)$ as
\begin{equation}
J_{\alpha}=\frac{c(1+z)^2}{4\pi} \sum_{n=2}^{n_{max}} f_{rec}(n) \int_z^{z_{max,n}} dz'\frac{\epsilon_{\alpha}(\nu'_n,z')}{H(z')}~,
\end{equation}
where the emission frequency is $\nu'_n= \nu_n(1+z')/(1+z)$,
$\nu_n=\nu_{LL}(1-n^{-2})$, $\nu_{LL}$ is the Lyman limit frequency and
$H(z)$ is the Hubble expansion rate.

 We use a simple model for the emissivity, based on a constant emission
per unit mass of collapsed halos above a minimum mass $M_{\rm min}$, at
the moment mass is added to them~\cite{Barkana:2004vb},

\begin{equation}
\epsilon_{\alpha}(\nu,z) = 
 \varepsilon(\nu) f_* \bar{n}_{b,0}  \frac{df_{coll}(z)}{dt}~,
\end{equation}
where $f_*$ is the fraction of baryons that form stars when new mass is
added to collapsed halos, and
$\bar{n}_{b,0}$ is the mean comoving baryon number density. We take $f_*=0.01$ following previous works on radiation-hydrodynamic simulations of high-redshift galaxies (e.g. \cite{Wise:2014vwa}) or on the comparison of the star formation rate density to the one derived from UV luminosity function measurements \cite{Lidz:2018fqo}. The fraction
of mass collapsed in halos which host star formation $f_{coll}$ is
\begin{equation}\label{hmf}
f_{coll}(z) = \frac{1}{\rho_m}
 \int_{M_{min}(z)}^{\infty} dM \, M \, \frac{d n}{dM}~,
\end{equation}
where $\frac{d n}{dM}$ is the halo mass function, computed here using
the Sheth-Tormen function~\cite{Sheth:1999mn,Sheth:1999su}. The minimum
mass $M_{\rm min}$ is fixed here to the virial mass corresponding to a
halo virial temperature of $T_{\rm vir}=10^4$ K. This neglects star
formation that can occur at lower virial temperatures via molecular
cooling, and assumes that most of the emissivity is due to stellar
populations formed when atomic hydrogen cooling of partially ionized
matter is already important. In any case, equation (\ref{hmf}) is only
a simple but reasonable model for the evolution of the emissivity from
first galaxies.
The radiation spectral distribution, $\varepsilon(\nu)$, is normalized
to $\int_{\nu_{\alpha}}^{\nu_{LL}} d\nu \varepsilon(\nu) = N_{\rm tot}
\simeq 9690$,
using the total number of photons emitted between the Ly$\alpha$ and the
Lyman limit, $N_{\rm tot}$, from the Population II (or low-metallicity)
model of \cite{Barkana:2004vb}. Given the uncertainty in the spefic spectral shape, we assume a emissivity proportional to $\nu^{-1}$,
resulting in the normalization $\varepsilon(\nu) = N_{tot}/{\rm ln}(4/3)\,
\nu^{-1}$. All in all, the Ly$\alpha$ flux is
\begin{equation}
\begin{split}
J_{\alpha}=&\frac{c(1+z)^3f_* \bar{n}_{b,0}N_{tot}}{4\pi\nu_\alpha ln(4/3)}  \times \\
& \times \sum_{n=2}^{n_{max}} f_{rec}(n) \frac{\nu_\alpha}{\nu_n} \left(f_{coll}(z)-f_{coll}(z_{max,n})\right)~,
\end{split}
\end{equation}
with $\nu_\alpha=\nu_2$ the Ly$\alpha$ frequency. This flux differs by less than 10\% from the one computed with the
more complex model of a piecewise power-law spectrum presented in
\cite{Barkana:2004vb}.

\subsection{Heating sources}

The kinetic temperature $T_K$ can be determined by solving the thermal evolution equation
\begin{equation}
\frac{d T_K}{dt} + 2 H T_K - \frac{2}{3} \frac{T_K}{\Delta}\frac{d \Delta}{dt}
 +\frac{T_K}{1+x_e}\frac{d x_e}{dt} = \frac{2  \mathcal{Q}}{3 \, n_b(1+x_e)}~,
\label{eq:eqT}
\end{equation}
where $x_e$ is the ionized fraction leftover from recombination
($x_e \sim 10^{-4}$), $n_b=\bar{n}_{b,0}(1+z)^3\Delta$ is the baryon number
density and $\mathcal{Q}$ is the total heating rate per unit volume.

The expected dominant heating mechanism is X-ray heating from astrophysical
sources (e.g. \cite{Furlanetto:2006tf}). We also account for two other
usually neglected model-independent heating sources: CMB photons causing
21-cm transitions and Ly$\alpha$ scatterings. Compton cooling is also taken
into account, although it is negligible due to the small ionized fraction
in the IGM prior to reionization.

\begin{itemize}

\item \textbf{21-cm heating}:

The heating rate due to absorption and emission of CMB photons by the
hyperfine levels was derived in \cite{Venumadhav:2018uwn} and can be written as
\begin{equation}
\mathcal{Q} |_{21} = \frac{3}{4} n_H x_{\rm HI} x_{CMB} A_{10} T_* \left( \frac{T_\gamma}{T_S} -1 \right),
\label{eq:heat21}
\end{equation}
where $n_H=\bar{n}_{H,0}(1+z)^3\Delta$ is the number density of hydrogen,
$\bar{n}_{H,0}$ is its comoving average, and
$x_{CMB} = (1-e^{-\tau_{21}})/\tau_{21} \simeq 1$, where the 21-cm optical
depth $\tau_{21}$ is small.

\item \textbf{Ly$\alpha$ heating}

Another heating source is Ly$\alpha$ scattering which, following
\cite{Chen:2003gc, Furlanetto:2006fs}, has a contribution from continuous and
injected photons (photons that redshift into the Ly$\alpha$ resonance from
the continuum, or are produced at the Ly$\alpha$ resonance after absorption
at a higher energy line, respectively) of
\begin{equation}
\mathcal{Q} |_{\rm Ly\alpha,k} = \frac{4 \pi H h\nu_\alpha \Delta\nu_D}{c} J_{\infty,k}I_k~,
\label{eq:heatLya}
\end{equation}
where $k$ stands for $k=c$, \textit{continuous}, or $k=i$, \textit{injected}, with $\Delta\nu_D =  \nu_\alpha \sqrt{2k_BT_K/(m_p c^2)}$ the Doppler broadening parameter.
The integrals $I_c$ and $I_i$ encode the details on the scattering effects and
depend on the temperature and the Gunn-Peterson optical depth $\tau_{GP}$ (the
optical depth of Ly$\alpha$ photons redshifting through the IGM). For these
quantities, we use the fits provided by \cite{Furlanetto:2006fs}. 

\item \textbf{X-ray heating}

The heating from astrophysical X-ray sources is accounted for with the
on-the-spot approximation (e.g.\cite{Furlanetto:2006tf}),
\begin{equation}
\mathcal{Q} |_{X} = \xi_{X}\,\epsilon_X\, f_* f_{\rm heat}\,
 \mu m_p n_b\, \frac{d f_{coll}}{dt}
\label{heatXray}
\end{equation}
where we compute the fraction of energy from X-rays that is converted to
heat in the IGM, $f_{\rm heat}$, using the fit 
from \cite{1985ApJ...298..268S}. The fiducial value for the luminosity of
the X-ray emission per unit star formation rate is $\epsilon_X=
3.4 \times 10^{40}$~erg s$^{-1} M_\odot^{-1}$~yr,
from \cite{Furlanetto:2006tf}, which is measured from local starbust
galaxies. The efficiency parameter $\xi_{X}$ accounts for deviations
from this fiducial value, so in the default model, $\xi_{\rm X}=1$.

\end{itemize}

\section{IGM density distribution}
\label{sec:sec2}

\begin{figure}[t]
\begin{center}
	\includegraphics[width=0.49\textwidth]{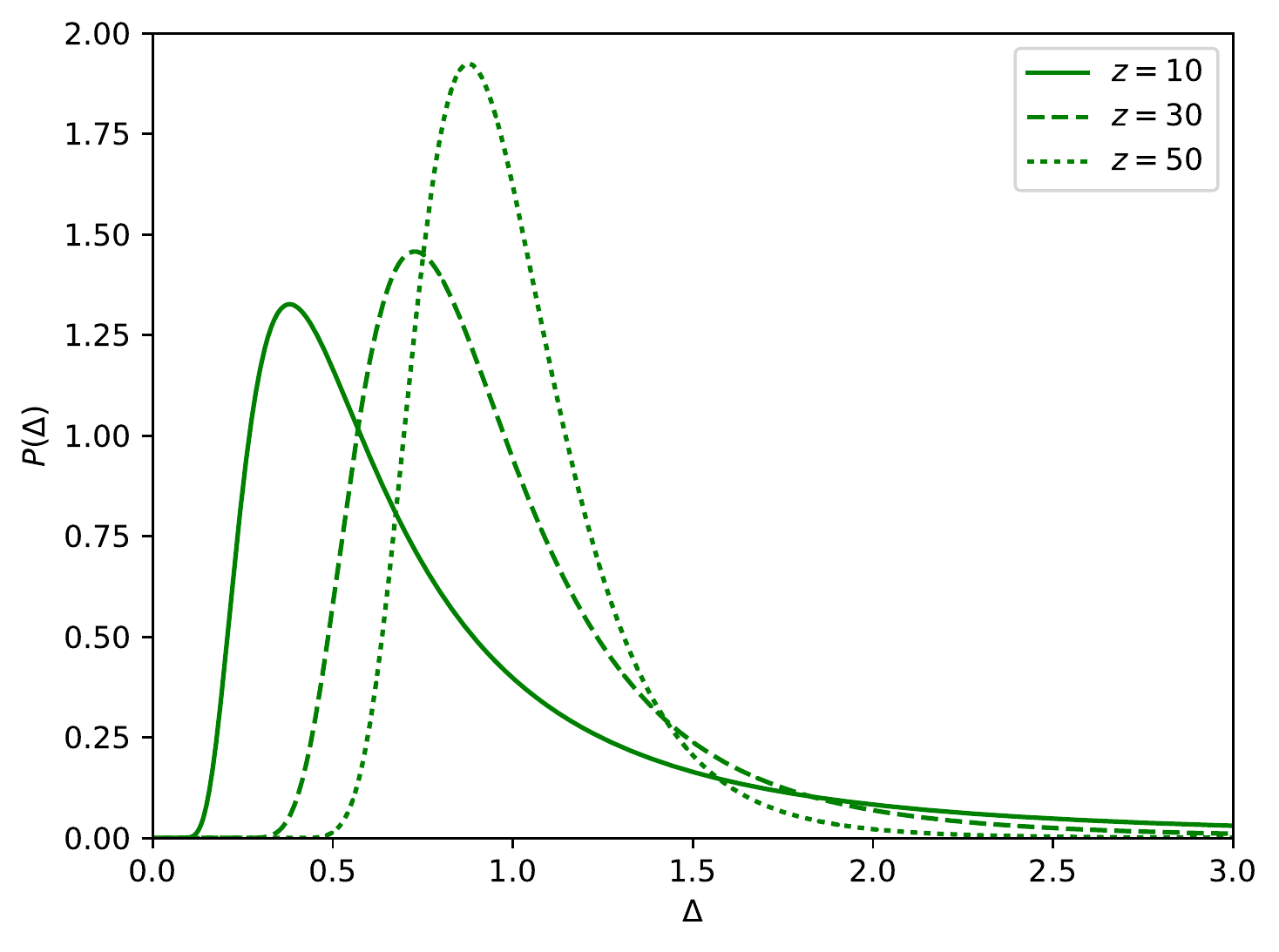}
	\caption{MHR00 probability distribution of the IGM density for
three different redshifts.}
	\label{fig:probs}
	\end{center}
\end{figure}

 The mean 21-cm absorption can be affected by the presence of spatial
density fluctuations in the atomic gas density of the IGM,
$\rho$. We define the non-linear probability
distribution function (PDF) of the gas density, $P(\Delta)$, where
$\Delta=1+\delta=\rho/\bar\rho$.
While the original linear distribution is described by a Gaussian,
the PDF becomes non-Gaussian when fluctuations at
the Jeans scale of the atomic IGM reach non-linearity.
The probability and mass normalization imply that:
\begin{equation}
\int_{0}^\infty d\Delta P(\Delta) = 1~, \; \; \; \int_{0}^\infty d\Delta \; \Delta P(\Delta) = 1~.
\label{eq:norm}
\end{equation}
In the following, we denote the average of any function $f$ over the density distribution as
\begin{equation}
\langle f \rangle = \int_{0}^\infty d\Delta \; f(\Delta) P(\Delta)~,
\end{equation}
and the second central moment of the distribution, or variance, as $\sigma_\delta^2 = \langle (\Delta-1)^2 \rangle = \langle \delta^2 \rangle$. We use $\overline{f}$ for the background values, when $\delta=0$, which, in general, will differ from the value of $\langle f \rangle$.

 As a fit to the non-linear density distribution,
the following probability distribution
(named in what follows as MHR00) was proposed to fit numerical simulation
results by \cite{MiraldaEscude:1998qs},
\begin{equation}
P_{\rm{MHR00}}(\Delta)=A\Delta^{-\beta}\exp\left[ -\frac{(\Delta^{-2/3}-C_0)^2}{2(2\delta_0/3)^2} \right],
\label{eq:MHR00}
\end{equation}
where $A$, $\beta$, $C_0$ and $\delta_0$ are free parameters constrained
by the normalization conditions in equations (\ref{eq:norm}). Fig. \ref{fig:probs} depicts the $\rm{MHR00}$ PDF for three different redshifts. This
distribution is motivated to describe the evolution of low-density regions,
where gravitational tides are usually small and test particles move away
from each other at roughly constant velocity. Therefore,
densities decrease as $\rho \sim t^{-3} \sim a^{-9/2}$ in a
matter dominated universe and consequently, $\Delta = \rho/\bar{\rho} \sim a^{-3/2}$. Since the
linear density field grows as $\delta_l \propto a$, we can write
$\Delta \sim \delta_l^{-3/2}$. Therefore, an initial Gaussian distribution
in $\delta_l$ gives rise to a Gaussian PDF in $\Delta^{-2/3}$, as in
eq.~(\ref{eq:MHR00}).
In the linear regime limit, when $\delta_0 \ll 1$, a Gaussian
distribution with dispersion $\delta_0$ and $C_0=1$ is recovered.
For high densities, $\Delta \gg 1$, the probability distribution tends to
a power-law, $P_{\rm{MHR00}} \propto \Delta^{-\beta}$. This is easily seen to
correspond to a power-law density profile in collapsed objects
$\Delta \propto r^{-3/(\beta-1)}$.
For our numerical model, we choose $\beta=5/2$ which
corresponds to isothermal halos with overdensity $\Delta \propto r^{-2}$.
This distribution was seen to provide a reasonable fit to the results
of numerical hydrodynamic simulations for the evolution of a photoionized
IGM, simulating the Ly$\alpha$ forest, in
Ref.~\cite{MiraldaEscude:1998qs}.

 For the dispersion of the PDF, we compute a filtered variance from the
linear power spectrum using a top-hat sphere of comoving radius $R_J$ as
filter,
\begin{equation}
\sigma_l^2(R)=\int \frac{d^3k}{2\pi}P(k)|W(kR_J)|^2 ~.
\label{eq:sigma}
\end{equation}
The natural smoothing length for the gas density is the Jeans length at the
temperature of the atomic medium, which sets the scale where the
collapse of the first halos with high gas overdensities is expected.
We use a top-hat filtering radius $R_J = \lambda_J/4$, with the Jeans
length $\lambda_J$ defined in eq. (\ref{jeans}),
fixing therefore $\delta_0=\sigma_l(R_J)$. This roughly corresponds to
the scale of the first non-linear collapse followed by the atomic gas,
because the top-hat diameter corresponds to half the full critical
wavelength for Jeans instability (the region where the density
perturbation is positive). The mass enclosed in a sphere of radius $R_J$
is also close to the Bonnor-Ebert critical gas mass for gravitational
instability \cite{Draine2011}.

 Finally, we set an upper cutoff to the distribution at
$\Delta_{\rm max}=10$, fixing the probability distribution to zero at
$\Delta > \Delta_{\rm max}$. This is reasonable when the
mass fraction that has collapsed on scales substantially larger than
$R_J$ is very small, because the atomic gas does not cool effectively
and behaves adiabatically, so it cannot collapse to high densities on
halos formed from scales comparable to $R_J$.
The values of the parameters $A$ and $C_0$ are then determined by the
normalization conditions of eq.~(\ref{eq:norm}).

Other choices for the PDF can be considered, such as the log-normal
distribution (e.g. \cite{Coles:1991if,Becker:2006qj}).
We have checked our results for this other distribution, finding no
strong dependence on the shape of the PDF of $\Delta$ as long as the
characteristic width of the distribution is kept fixed.

\section{Impact of IGM density fluctuations on the 21-cm signal}
\label{sec:sec3}

 We now present how the 21-cm absorption is affected by the presence of
density fluctuations in the atomic IGM, in various models for the
heating sources described above and for the Ly$\alpha$ radiation
coupling the spin and kinetic temperatures. We have developed our own numerical implementation as a Python code~\footnote{The Python code
used for the calculations is made publicly available at
\url{https://github.com/PabloVD/21cmSolver}.}
which solves the thermal evolution accounting for density fluctuations and
the various heating terms. Other codes are available in the literature
to compute the 21-cm global signature. For instance, the program 21cmFAST
\cite{Mesinger:2010ne} computes the space-dependent 21-cm signal from a 3D data cube of
the density field, therefore accounting for the impact of the density
fluctuations automatically. Our procedure, on the other hand, accounts for
density fluctuations analytically to compute the global signal. Another
often used code is ARES \cite{2014MNRAS.443.1211M}, which computes the global signal with
a more realistic radiative transfer calculation of the X-ray heating compared
to our on-the-spot approximation of equation \ref{heatXray}, but does not take into
account density fluctuations. 

\subsection{Strong Ly$\alpha$ coupling in adiabatic cooling regime}
\label{strongnoheat}

\begin{figure}[th]
\begin{center}
	\includegraphics[width=0.5\textwidth]{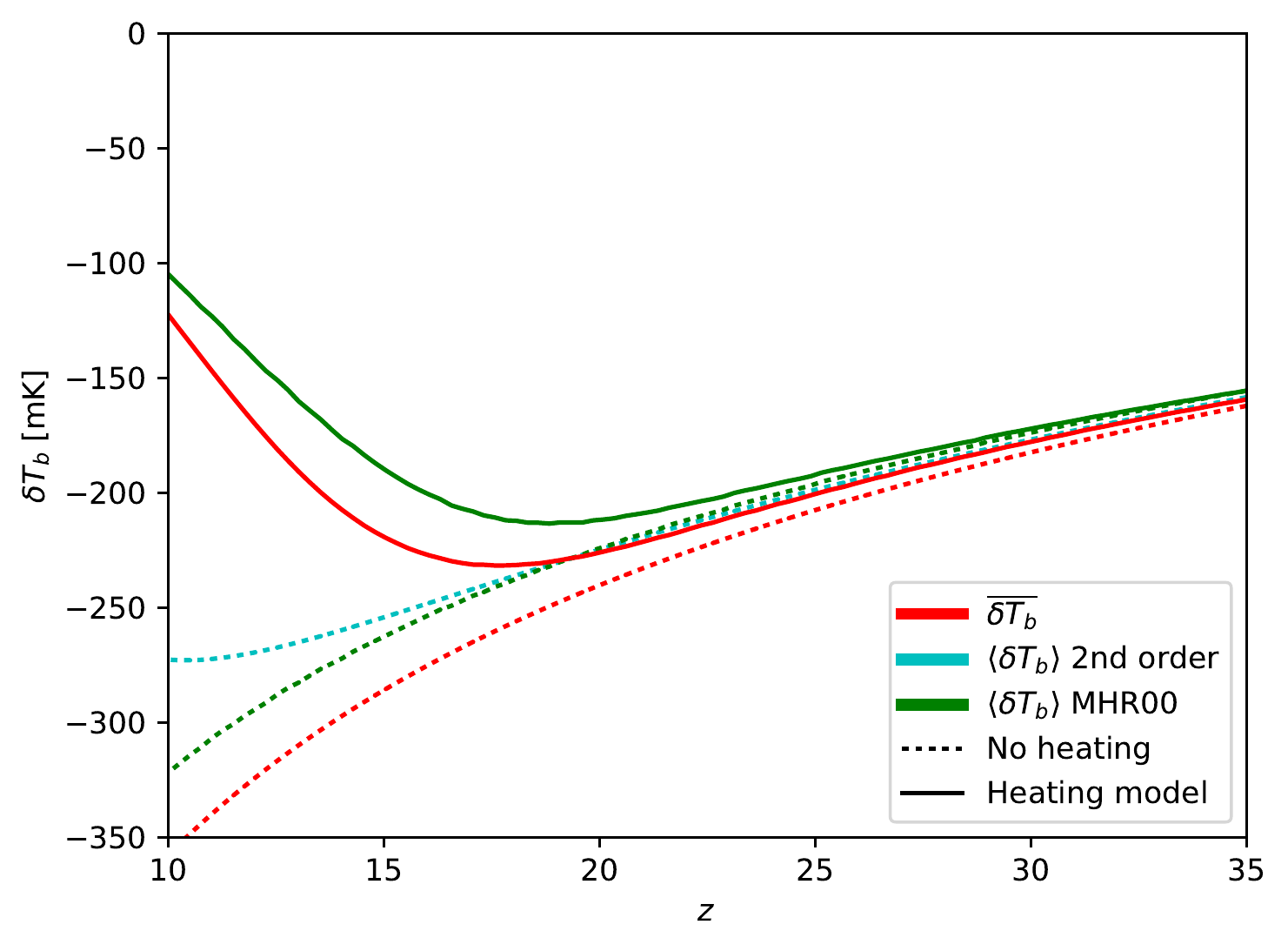}
	\caption{21-cm global brightness temperature in the strong
Ly$\alpha$ coupling regime at the mean density $\Delta=1$ ({\it red}),
and the mean value $\langle \delta T_b \rangle$ accounting for density
fluctuations using the second-order approximation of
eq.~(\ref{dTboverdense}) ({\it cyan}), or the MHR00 distribution
({\it green}). Dotted lines are for adiabatic cooling while solid lines
include 21-cm and Ly$\alpha$ heating terms in our model.}
	\label{fig:dTbAbs}
	\end{center}
\end{figure}

 As a first illustrative case, we consider the regime of perfect spin
and kinetic temperature coupling by Ly$\alpha$ scattering,
while neglecting all heating terms. Therefore, $x_{\alpha} \gg 1$ and
$T_S=T_K$ (see eqs. \ref{eq:dTbdelta} and \ref{eq:Tspin}). This
scenario represents the case of a very fast transition to strong
coupling before there is any time for substantial heating, so the
maximum 21-cm absorption is produced. If the gas cools adiabatically
from high redshift in the absence of any heating, the kinetic
temperature of the monoatomic gas is related to density as
$T_K = \overline{T}_{ad} \Delta^{2/3}$ (see eq.~\ref{eq:eqT}), where
$\overline{T}_{ad} \propto a^{-2}$ is the kinetic temperature at $\Delta=1$, or $\delta=0$.
The mean background 21-cm emission is
\begin{equation}
\overline{\delta T_b} = T_0 \left(1- \frac{T_\gamma}{\overline{T}_{ad}} \right)~.
\label{eq:dTbm}
\end{equation}

The 21-cm background signal is obtained by averaging equation
(\ref{eq:dTbdelta}) over all densities with the strong coupling
assumption, $T_S(\Delta)=T_K(\Delta)$:
\begin{align}
\langle \delta T_b \rangle &= T_0 \langle \Delta
 \left(1- \frac{T_\gamma}{\overline{T}_{ad} \Delta^{2/3}} \right) \rangle =
 T_0 \left( 1 - \frac{T_\gamma}{\overline{T}_{ad}} \langle \Delta^{1/3} \rangle \right)  \\
 &= \overline{\delta T_b} + T_0\frac{T_\gamma}{\overline{T}_{ad}}
 (1- \langle \Delta^{1/3} \rangle) \label{dTboverdense1}\\
 & = \overline{\delta T_b} + \frac{1}{9} T_0\frac{T_\gamma}{\overline{T}_{ad}}
  \sigma_\delta^2 + O(\delta^3)~,
\label{dTboverdense}
\end{align}
where the last equation is valid up to second order in $\delta$, and
results from expanding $\langle (1+\delta)^{1/3} \rangle$ in moments of
the $\delta$ distribution. The impact of density fluctuations is a
positive quantity at the second order, added to the global signal,
diminishing therefore the maximum possible absorption.

 Figure~\ref{fig:dTbAbs} compares the mean absorption 21-cm signal
without density fluctuations $\overline{\delta T_b}$ ({\it dotted red
line}), to $\langle \delta T_b \rangle$ computed from the second order
approximation of eq.~(\ref{dTboverdense}) ({\it dotted cyan}), and to the same mean signal computed from the full
MHR00 PDF ({\it dotted green}). In the latter case, we use also
eq.~(\ref{dTboverdense1}), which is highly accurate in the absence of
the heating terms discussed in Section 2 because the normalized density
$\Delta$ varies appreciably only when Compton heating of the atomic IGM
by the CMB is already small. As expected, density fluctuations imply a
lower absorption amplitude by $\sim$ 10\%, and the second order
approximation is close to the exact PDF calculation, deviating
substantially only at $z\lesssim 15$.

\subsection{Inclusion of 21-cm and Ly$\alpha$ heating}
\label{strongheat}

 We now use the simple model of heating sources presented in Section 2
to investigate the impact of the small-scale density fluctuations in the
atomic IGM on the 21-cm absorption evolution as the IGM temperature is
increased. As an illutrasting exercise, we first maintain the assumption
of a perfect coupling of the spin and kinetic temperatures (therefore
ignoring the Ly$\alpha$ emissivity discussed in section 2 and assuming
the limit of large $x_\alpha$ in eq. \ref{eq:Tspin}), but include the heating
derived from the Ly$\alpha$ flux and the CMB heating. This is of course
not a physically consistent model but it simply allows us to visualize
the impact of heating separately from that of the spin and kinetic
temperature coupling. The red solid line in Fig.~\ref{fig:dTbAbs} is
the result without density fluctuations. For this model, the maximum
absorption is reached at $z_{\rm max} \simeq 17$.

  We then include the density fluctuations using the model MHR00, using
the following simplification: we solve equation (\ref{eq:eqT}) for a
grid of many values of $\Delta$, neglecting the term $d\Delta/dt$. We
start the calculation at $z_i=35$ (at an epoch before there is any
significant heating), assuming the initial relation $T(z_i,\Delta)=
\overline{T}_{ad}(z_i) \Delta^{2/3}$, we evolve the temperature at each $\Delta$ as a
function of redshift, and then we compute the mean absorption using
the exact equation,
\begin{equation} \label{eq:tbg}
\langle \delta T_b \rangle =T_0 \left( 1- \langle\Delta\, {T_\gamma \over T_S(\Delta) }\rangle \right)~.
\end{equation}
Our justification for neglecting the term $d\Delta/dt$ is that the
heating sources increase rapidly with time, so that when evaluating
the mean absorption at any redshift $z$, most of the heating affecting
$T_K(z)$ has occurred over a brief time just before redshift $z$.
Taking into account the term $d\Delta/dt$ is complex, because at
second order the evolution of $\Delta$ is no longer local and one has
to compute an average over evolutionary histories of volume elements
in the IGM. Furthermore, at high $\Delta$ shock-heating will inevitably
occur in the non-linear regime. While a number of approximations may be
considered for this term, the results would not change dramatically
because of the rapid rise of heating sources from the collapse of
high-$\sigma$ halos.

 The result when we include only 21-cm and Ly$\alpha$ heating (with
no X-rays) is shown as the green solid line in Figure~\ref{fig:dTbAbs},
showing the same decrease of the absorption amplitude ($\sim 30$ mK at
$z\lesssim 18$) compared to the red solid line, and a shift of the
redshift of maximum absorption to
$z_{\rm max}\simeq 19$. There is therefore a substantial and measurable
impact of the IGM small-scale density fluctuations on the global
21-cm absorption history, although the details of the variation of
$\overline{ \delta T_b }$ with redshift depend on the heating history.

\begin{figure}[th]
\begin{center}
	\includegraphics[width=0.5\textwidth]{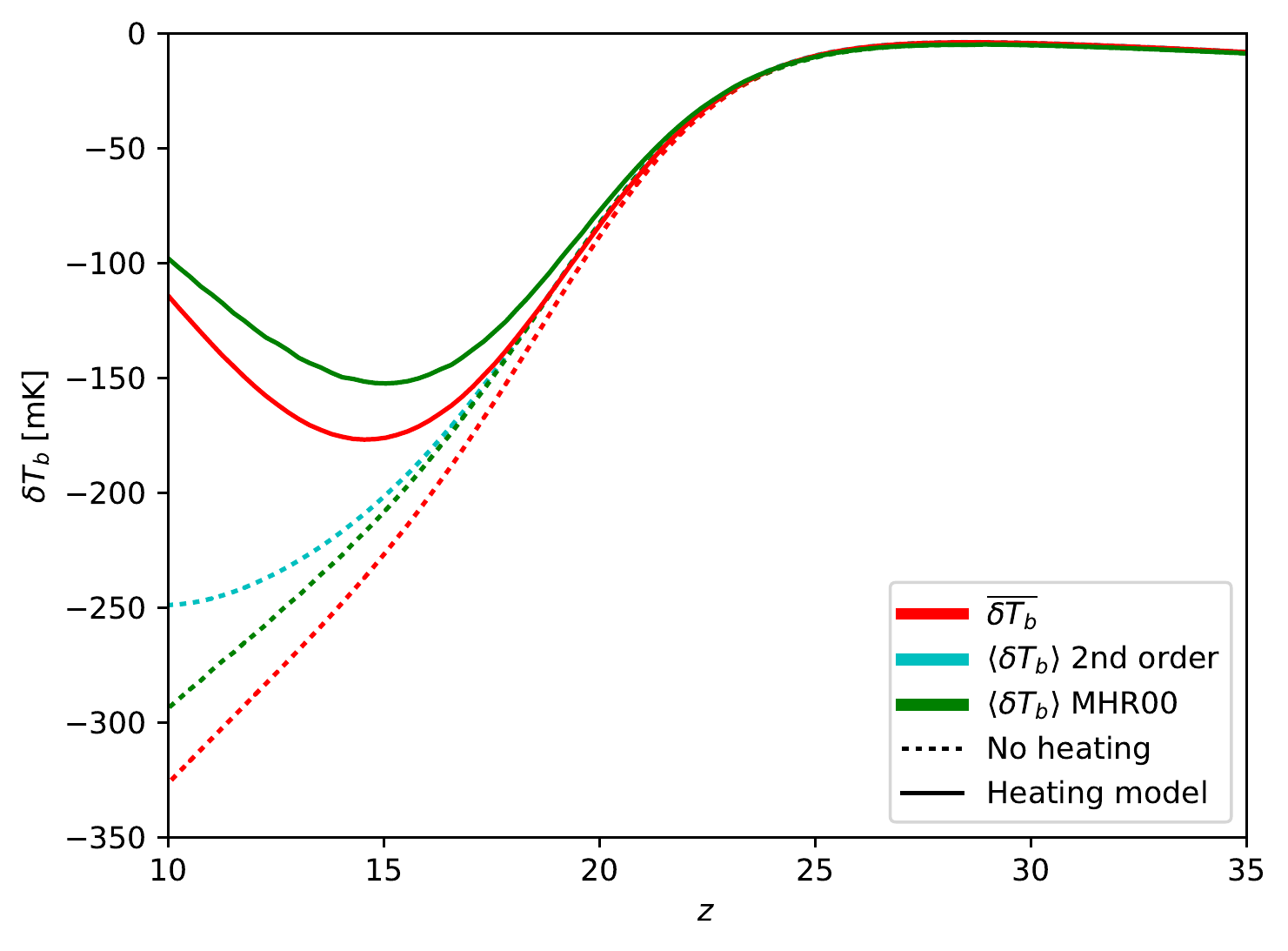}
	\caption{Same as Fig. \ref{fig:dTbAbs} but including the correct
Ly$\alpha$ coupling to calculate the evolution of the spin temperature. }
	\label{fig:dTbAbsCoup}
	\end{center}
\end{figure}

 Next, we include the correct Ly$\alpha$ coupling from the model
described in Section 2 to calculate the evolution of the spin
temperature. In the adiabatic cooling case, we can write analitically the background 21-cm brightness temperature as
\begin{equation} \label{eq:tbgc}
\overline{\delta T_b}= T_0 { \, x_{tot}\over 1+x_{tot} }
 \left(1- {T_\gamma \over \overline{T}_{ad}}\right)~.
\end{equation}
where $x_{tot}=x_\alpha + x_c$. When the Ly$\alpha$ flux, assumed to be homogeneous and independent of the density $\Delta$, dominates the coupling in
eq.~(\ref{eq:Tspin}), then $x_{tot} \simeq x_\alpha$. As before, the mean 21-cm brightness can be approximated to second
order in $\delta$, yielding
\begin{equation}
\langle \delta T_b \rangle \simeq \overline{\delta T_b} +
 \frac{1}{9} T_0\frac{T_\gamma}{\overline{T}_{ad}}
 \frac{x_{tot}}{1+x_{tot}} \sigma^2_\delta~.
\label{dTboverdensexalpha}
\end{equation}

The results using this formula, together with the numerical integration using the full PDF, are depicted with dotted lines in
 Fig.~\ref{fig:dTbAbsCoup}. The dotted lines (no heating)
now show a rapid increase of the absorption signal with time because of
the rising Ly$\alpha$ intensity, with coupling becoming effective only
at $z\lesssim 20$. The inclusion of heating (solid lines in Fig. \ref{fig:dTbAbsCoup}) now produces a reduced
absorption with a maximum delayed to $z_{\rm max}\simeq 15$, because of
the imperfect coupling, and the presence of density fluctuations also
cause as before a further reduced absorption amplitude, of the order of $\sim 15\%$ respect to the background case.

\subsection{General Ly$\alpha$ coupling with heating sources}
\label{sec:secx}

\begin{figure}[t]
\begin{center}
	\includegraphics[width=0.5\textwidth]{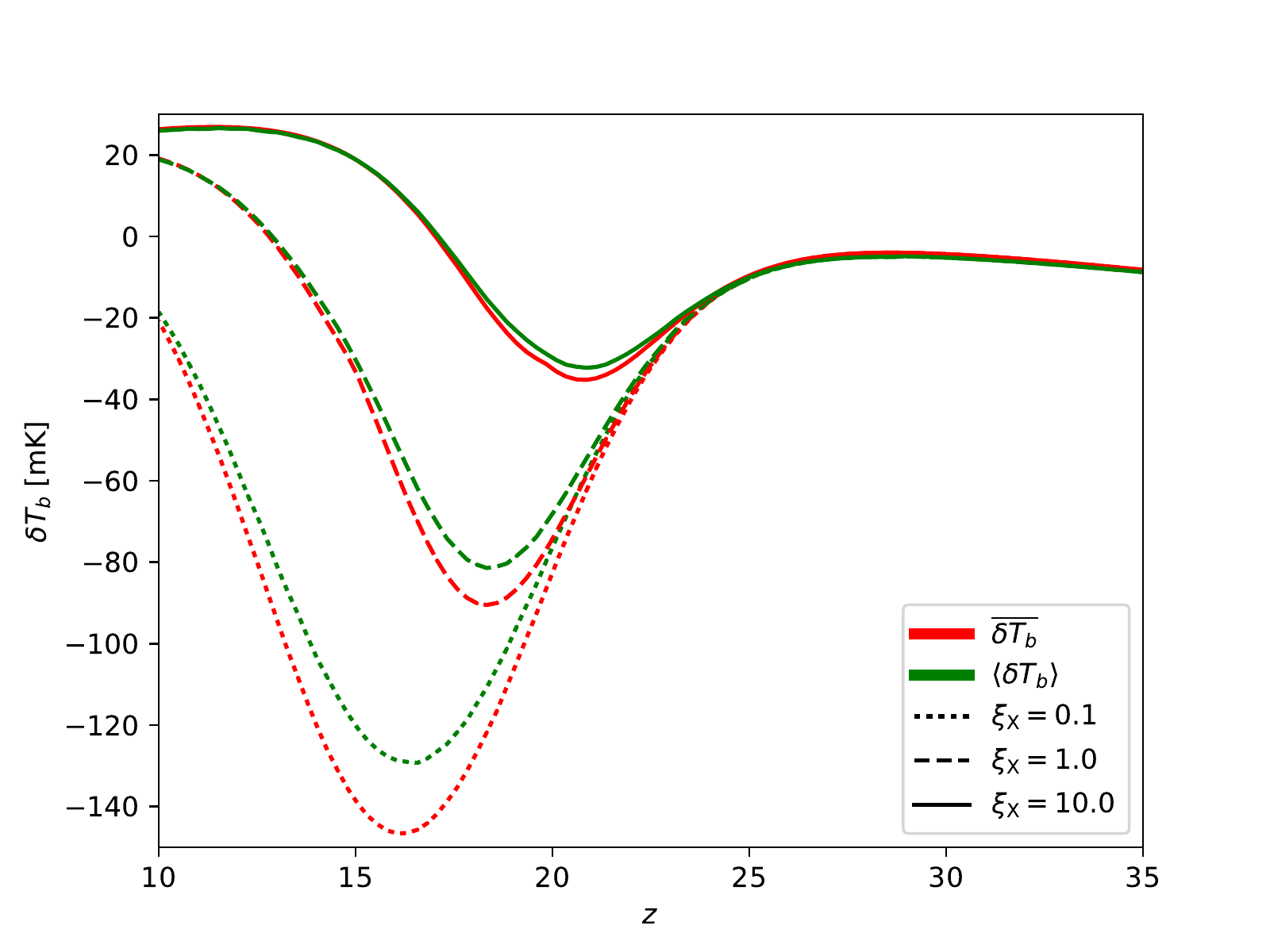}
	\caption{Same as Fig. \ref{fig:dTbAbsCoup} but including X-ray
heating for three different galaxy emissivities, where $\xi_X=1$
corresponds to X-ray emission per unit star formation similar to local
starbursts.}
	\label{fig:dTbAbsXray}
	\end{center}
\end{figure}

\begin{figure}[th]
\begin{center}
	\includegraphics[width=0.5\textwidth]{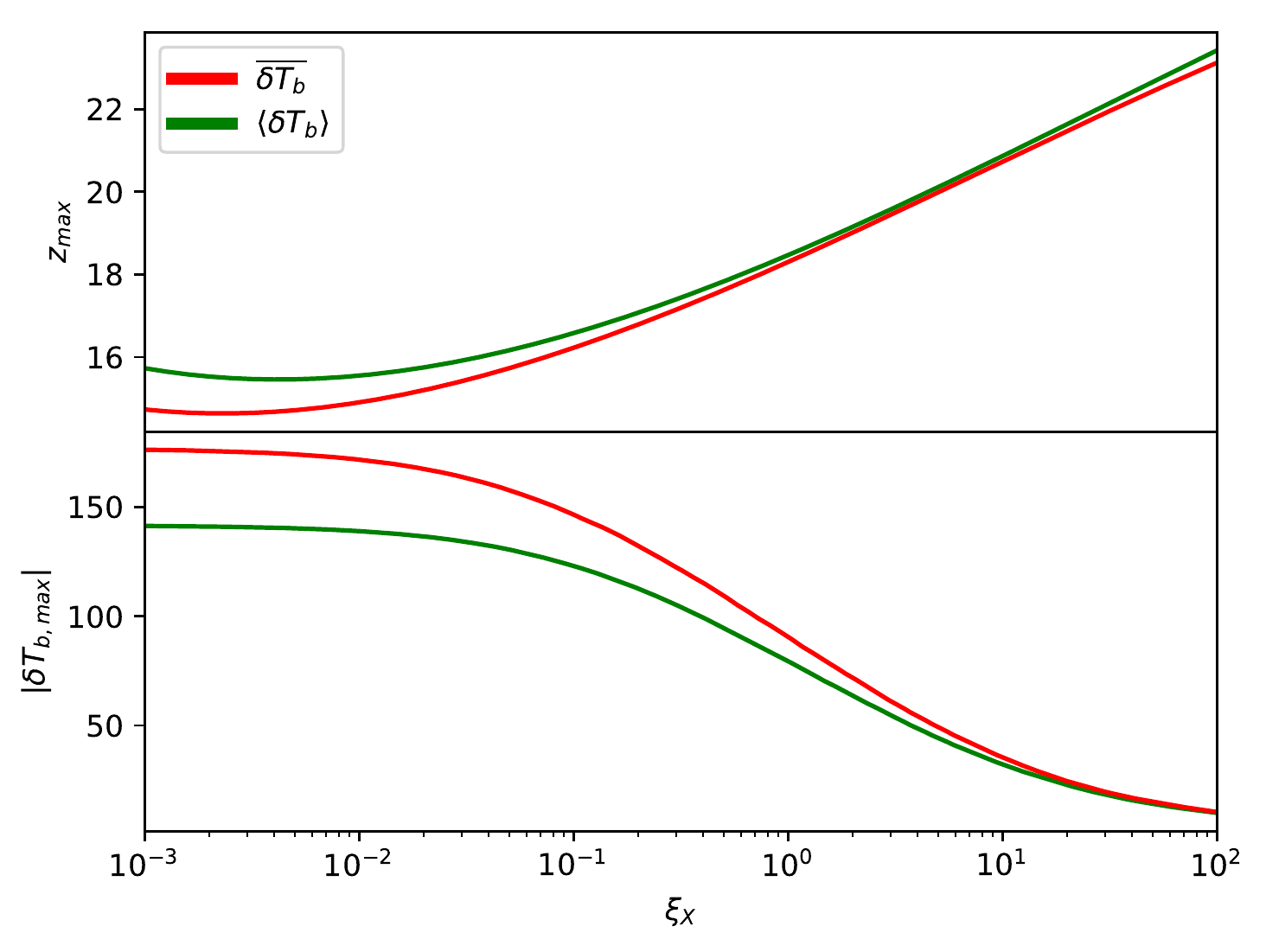}
	\caption{Maximum absorption $| \delta T_b| $ and redshift
when it is reached, as a function of X-ray emission efficiency.}
	\label{fig:TbAbsXray}
	\end{center}
\end{figure}

 Finally, Fig.~\ref{fig:dTbAbsXray} shows the calculated mean 21-cm
brightness when X-ray heating is included according to
eq.~(\ref{heatXray}), assuming the three different values of the X-ray
efficiency $\xi_X = (0.1, 1, 10)$. For an X-ray emissivity comparable to
local starbursts, the maximum 21-cm absorption is reduced from
$\sim 140$ mK to $\sim 80$ mK, with a shift to a higher maximum redshift
$z_{\rm max}\simeq 18$. With $\xi_X=0.1$ there is only a small reduction
in the maximum 21-cm absorption, while with $\xi_X=10$ the absorption is
greatly reduced to a brightness temperature comparable to the 21-cm
emission that is already induced by $z\simeq 15$. Of course, the
detailed epoch and value of the maximum 21-cm absorption depends on
the model we have assumed for the Ly$\alpha$ emission rate as a result
of star formation in the earliest, smallest galaxies to form, but the
results of this model illustrate how the observed 21-cm signal depends
on the model and the impact of the small-scale density fluctuations in
the atomic IGM. This is shown more clearly in Figure~\ref{fig:TbAbsXray},
where the maximum 21-cm absorption $\delta T_b$ (lower panel) and the
redshift $z_{\rm max}$ at which this minimum occurs (upper panel) are
plotted as a function of the X-ray efficiency $\xi_X$, when all three
heating sources (21-cm, Ly$\alpha$ and X-ray) are included. As before,
the red and green curves show the evolution at $\Delta=1$, and when the
density distribution is numerically included as described above. For
$\xi_X<10^{-2}$, X-ray heating is negligible and the maximum possible
21-cm absorption is reached, decreased (in absolute value) by $\sim 30$ mK
by the presence of density fluctuations, while for high X-ray emissivity,
the absorption amplitude becomes increasingly small and is reached at an
increasing redshift.

\section{Conclusions}
\label{sec:sec4}

 The properties of the earliest star formation period in the Universe are
largely unknown. Understanding this epoch requires detecting matter
density fluctuations on tiny scales compared to the galaxy-scale
distribution we measure at low redshift, smaller also than the $1$~Mpc
scales that are probed by the Ly$\alpha$ forest in the spectra of
distant quasars. During these primitive stages, the interplay between
CMB photons, ultraviolet radiation and atomic hydrogen in the IGM leads
to the 21-cm absorption feature in the CMB spectrum. Understanding the
depth and redshift location of this absorption feature is crucial to
unravel the formation of stars in the first haloes and, ultimately, to
probe the nature of dark matter from the small-scale power spectrum.
From the observational perspective, the field of 21-cm cosmology is
entering a promising era of new data with expectations of first
detections in the near future. While numerical simulations are at hand,
analytical descriptions of the impact of small-scale density
fluctuations in the IGM on the 21-cm absorption profile were missing in
the literature. This has been the major goal of this paper: we have
discussed how these density fluctuations will significantly reduce the
amplitude of the 21-cm absorption peak, and how this absorption
reduction depends on the heating sources and the Ly$\alpha$ intensity
that determines the coupling of spin and kinetic temperatures.
This reduction is not highly dependent on the modeling of the shape of
the IGM density distribution, but the expected increase in the
distribution breadth due to gravitational evolution ends up producing
a nearly constant absorption temperature increase of $\sim 30$ mK at
redshifts below the maximum 21-cm absorption, $z < z_{\rm max}$. The
shape of the 21-cm absorption dip is highly uncertain, mostly due to
the unknown X-ray emission from the first star-forming dwarf galaxies
and the Ly$\alpha$ emission determining coupling, but we have shown that
the IGM density distribution modifies this absorption dip in a
substantial way in any case.

\section{Acknowledgments}
OM and PVD are supported by the Spanish grants FPA2017-85985-P and
SEV-2014-0398 of the MINECO, by PROMETEO/2019/083 and by the
European Union Horizon 2020 research and innovation program
(grant agreements No. 690575 and 67489). JM was supported in part by
Spanish grants AYA2015-71091-P and MDM-2014-0369.

\bibliography{biblio}

\end{document}